# Observation of an Unusual Colossal Anisotropic Magnetoresistance Effect in an Antiferromagnetic Semiconductor

*Huali Yang, Qing Liu, Zhaoliang Liao, Liang Si, Peiheng Jiang, Xiaolei Liu, Yanfeng Guo, Junjie Yin, Meng Wang, Zhigao Sheng, Yuxin Zhao, Zhiming Wang, Zhicheng Zhong\*, Run-Wei Li\**

Dr. Huali Yang, Qing Liu, Dr. Liang Si, Dr. Peiheng Jiang, Prof. Zhiming Wang, Prof. Zhicheng Zhong, Prof. Run-Wei Li
CAS Key Laboratory of Magnetic Materials and Devices & Zhejiang Province Key Laboratory of Magnetic Materials and Application Technology, Ningbo Institute of Materials Technology and Engineering, Chinese Academy of Sciences, Ningbo 315201, China
E-mails: zhong@nimte.ac.cn (Z.Z.); runweili@nimte.ac.cn (R.-W.L.)

Prof. Zhaoliang Liao
National Synchrotron Radiation Laboratory, University of Science and Technology of China, Hefei 230026, China

Xiaolei Liu, Prof. Yanfeng Guo
School of Physical Science and Technology, ShanghaiTech University, Shanghai 201210, China

Prof. Zhiming Wang, Prof. Zhicheng Zhong, Prof. Run-Wei Li
University of Chinese Academy of Sciences, Beijing 100049, China

Junjie Yin, Prof. Meng Wang
School of Physics, Sun Yat-Sen University, Guangzhou 510275, China

Prof. Zhigao Sheng
Anhui Key Laboratory of Condensed Matter Physics at Extreme Conditions, High Magnetic Field Laboratory, Chinese Academy of Sciences, Hefei, 230031 P. R. China

Qing Liu, Prof. Yuxin Zhao
National Laboratory of Solid State Microstructures and Department of Physics, Nanjing University, Nanjing 210093, China

Prof. Yuxin Zhao
Collaborative Innovation Center of Advanced Microstructures, Nanjing University, Nanjing 210093, China

**Searching for novel antiferromagnetic materials with large magnetotransport response is highly demanded for constructing future spintronic devices with high stability, fast switching speed, and high density. Here we report a colossal anisotropic magnetoresistance effect in an antiferromagnetic binary compound with layered structure — rare-earth dichalcogenide EuTe$_2$. The AMR reaches 40000%, which is 4 orders of magnitude larger than that in conventional antiferromagnetic alloys. Combined magnetization, resistivity, and theoretical analysis reveal that the colossal AMR effect is attributed to a novel mechanism of vector-field tunable band structure, rather than the conventional spin-orbit coupling mechanism. Moreover, it is revealed that the strong hybridization between orbitals of Eu-layer with localized spin and Te-layer with itinerant carriers is extremely important for the large AMR effect. Our results suggest a new direction towards exploring AFM materials with prominent magnetotransport properties, which creates an unprecedented opportunity for AFM spintronics applications.**

In antiferromagnets, the magnetic moments are aligned antiparallelly, giving rise to zero net moment. Although antiferromagnets have been discovered as the second basic type of magnetic materials as early as the 1930s, for quite a long period, they are considered impervious to external magnetic field and are mainly used inertly as pinning of the ferromagnetic layer in modern spintronics technology.[1] In recent years, it has been found that antiferromagnets are promising candidates for constructing fast-switching and high-density spintronic devices due to their ultra-fast spin dynamics (THz range) and vanishing stray field.[2,3] To make antiferromagnetic electronics feasible, a full electrical write and read strategy is demanded. Although encouraging progress has been made in the electrical manipulation and readout of antiferromagnetic order,[4-10] the small readout signal obstacles its real applications.

Anisotropic magnetoresistance (AMR) is a promising and widely investigated scheme to read out the antiferromagnetic state.[7-10] In analog to ferromagnets, the AMR of antiferromagnets mainly arises from the anisotropic spin dependent scattering due to spin-orbit coupling (SOC).[11] However, the AMR value are restricted to a few percent in most antiferromagnets,[7-10,12-14] which is incompatible with the requirement of keeping high signal-to-noise ratio at increasingly shortened readout time scale. Moreover, the influence of SOC on the electronic band structure is too complex to give a ready solution to obtaining a large AMR ratio. Here we take new perspective of AMR in antiferromagnets by considering their anisotropic response to external magnetic field.[15] For field along the easy axis, the antiferromagnetic spins remain collinear until the field is larger than the spin-flop field, the collinear spins preserve the twofold Kramers degeneracy inherent of antiferromagnets. However, applying field along the hard axis drives spin-canting, the net magnetic moment induced by spin-canting may induce band reconstruction by removing the twofold Kramers degeneracy.[16] As a result, the anisotropic response of band structure to vector field shall contribute to an unconventional AMR. To produce prominent AMR, the spins need be susceptible to external field and the band structure change should be prominent particularly near the Fermi surface. As we know, materials with localized spin moments are generally much more susceptible to external field than that with itinerant moments.[17] Unfortunately, electrons with localized spins usually lie deeply below the Fermi surface and can hardly contribute to transport properties. Hybridizing the orbital of localized spin with itinerant electron is a rational route to use localized spin for optimal magnetic response and the itinerant electrons near Fermi surface to manifest desired transport properties.

In this communication, we report the striking observation of a colossal AMR effect in the rare-earth dichalcogenide $EuTe_2$, an antiferromagnetic semiconductor with layered-structure. We performed combined experimental and theoretical studies and demonstrate that the strong hybridization between the Eu-layer with localized spin and the Te-layer with itinerant carriers contributes to a notable vector-field dependent band reconstruction, resulting in a colossal AMR of more than 40000%. These results demonstrate a promising route towards exploring antiferromagnets with prominent magnetotransport anisotropy and create an unprecedented opportunity for the technologically appealing AFM spintronics applications.

Single crystalline EuTe$_2$ samples were grown by self-flux methods and the high quality of the samples was characterized by x-ray diffraction (**Figure S1**). The EuTe$_2$ crystalizes in a tetragonal structure, in which Eu and Te atomic layers are alternately stacked along the *c* axis (**Figure 1a**). Figure 1b shows the temperature dependent magnetization (*M-T*) measured at 1 kOe after zero field cooling. The *M-T* curve shows typical AFM characteristic, and Curie-Weiss fitting of the magnetic susceptibility reveals that the effective moment is very close to the Eu atomic moment, indicating that the magnetic moment is mainly contributed by Eu.[18] The paramagnetic (PM) to AFM transition temperature ($T_N$) is determined to be 11.2 K. Below $T_N$, the magnetization decreases to near zero for *H // c*, whereas it is almost constant for *H // ab*. This fact indicates that the AFM easy axis is along *c* axis. Figure 1c illustrates the isothermal magnetization versus magnetic field (*M-H*) curves at 2 K for field along *c* axis and *ab* plane, respectively. For field parallel to easy axis (*H // c*), a spin-flop transition occurs at around 30 kOe. When the field is parallel to hard axis (*H // ab*), as a result of the spin-canting, the magnetization gradually increases with the magnetic field. At *H* > 100 kOe, the AFM spins are forced to align along the magnetic field direction, leading to a saturated magnetization moment of ~ 6.95 $\mu_B$ per unit cell.[18] This value is very close to the Eu atomic moment (7.0 $\mu_B$), consistent with the localized nature of Eu magnetic moment. Classic Monte Carlo simulation with Heisenberg model has been employed to simulate the *M-T* and *M-H* curves. The Heisenberg model with AFM exchange interaction is found to be in good agreement with experimental results,[19] further confirming the AFM characteristic of EuTe$_2$.

Now we turn to investigate the transport behavior, as well as its response to external fields. At zero field, the resistivity of EuTe$_2$ increases with reducing temperature, exhibiting an insulating behavior with a thermal-activation energy of less than 1 meV below $T_N$ (**Figure S4**).

Upon applying a magnetic field, the transport property is found to strongly depend on both field strength and field direction (Figure 1d). For example, when $H // c$ and $H = 25$ kOe (noted as $H_c = 25$ kOe for simplicity), the system experiences anomalous two phase transitions with one from insulator to metal at 6 K and the other from metal to insulator at 18 K. Quite interestingly, the resistivity changed by about two orders of magnitude at both transitions. With $H_c = 40$ kOe, only one transition from metal to insulator at ~ 32 K is observed in our measured temperature range of 2-100 K. Remarkably, the system exhibits dramatically different behavior for $H // ab$. In the case of $H_{ab} = 25$ kOe, EuTe$_2$ is metallic rather than semiconducting at low temperature, and the MIT occurs at 15 K, which is lower than that for $H // c$. At $H_{ab} = 40$ kOe, the MIT temperature is ~ 28 K, which is also lower than that for $H // c$. As can be seen from Figure 1D, the field direction dependent MIT exhibits different manner for temperature below and above $T_N$. To better illustrate this, we performed isothermal magnetoresistance (MR) measurements below and above $T_N$, respectively (Figure 1e). At 2 K ($< T_N$), the resistivity decreases with field, and steeply drops by almost two orders of magnitude at a critical field strength ($H_{cr}$), which corresponds to the MIT. The critical field strength ($H_{cr}$) for $H // c$ is ~ 30 kOe, which is much larger than that for $H // ab$ (~ 20 kOe). In contrast, at $T = 15$ K ($> T_N$), the critical field for MIT with $H // c$ is smaller than that with $H // ab$.

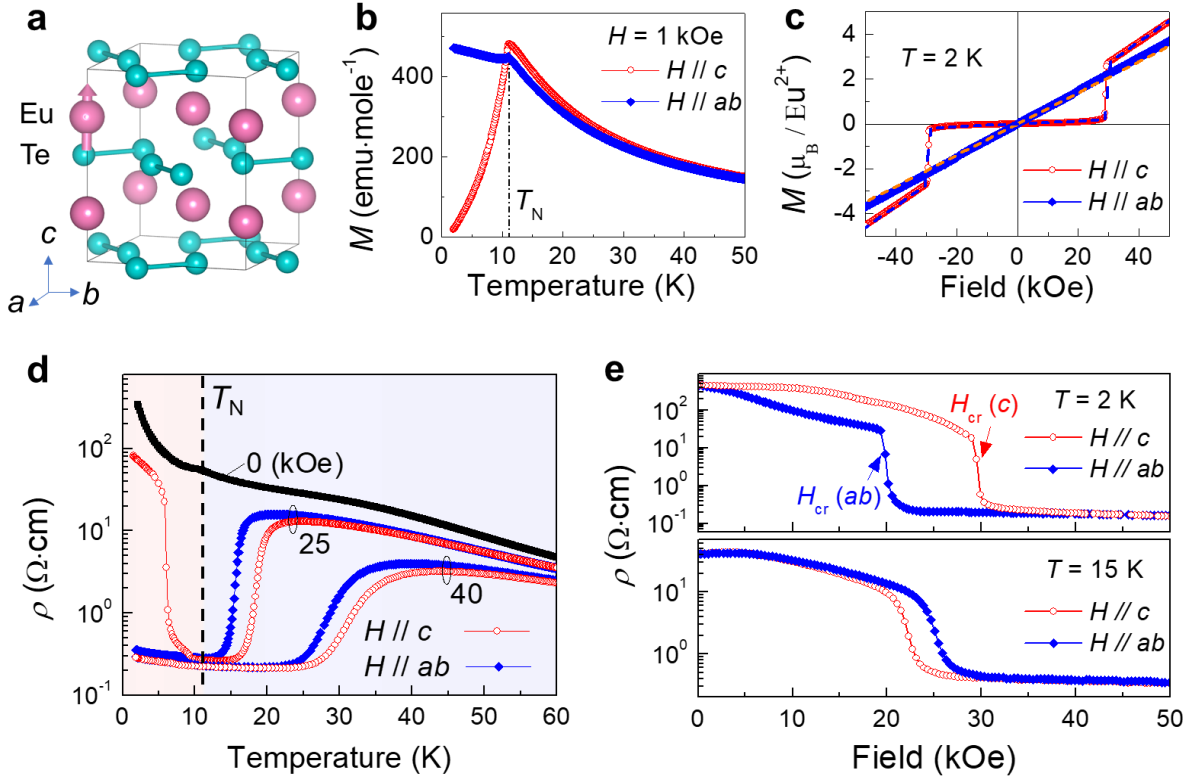

**Figure 1.** Magnetic and transport behavior of the EuTe$_2$ single crystal. a) Ball and stick model of the layered crystal structure of EuTe$_2$ unit cell. The Te atoms form dimers in the basal plane, and each Eu has eight nearest neighbor Te atoms. b) Temperature dependent magnetization (*M-T*) curves at 1 kOe for *H // c* and *H // ab* measured during warming process after zero-field cooling from 300 K to 2 K. c) Isothermal magnetizations versus magnetic field (*M-H*) curve with *H* along the *c* axis and *ab* plane at 2 K. The simulated curves are indicated by dashed lines. d) Temperature dependent resistivity with current *I // c* and *H* along *c* axis and *ab* plane, respectively. e) Isothermal magnetoresistance and field dependent magnetization for *H // c* and *H // ab* at 2 K and 15 K, respectively.

Due to the vector field dependent MIT, a large AMR is expected. **Figure 2a** depicts the AMR at 2 K as a function of $\theta$, which denotes the angle between *H* and principal axis *c*. The AMR exhibits a periodic behavior with uniaxial symmetry and very strong field dependence. For *H* < 19.5 kOe, the resistivity varies smoothly with $\theta$, the AMR [defined as ($\rho_{H // c}$ - $\rho_{H // ab}$) / $\rho_{H // ab}$] reaches ~6% under 1 kOe and amounts ~280% under 10 kOe. At 19.5 kOe < *H* < 29.5 kOe, the magnetoresistance shows an additional drop at specific critical angles ($\theta_c$). For example, at *H* = 19.7 kOe, the resistance drop appears when $\theta$ is near 0° and 180° (*H // ab*), and $\theta_c$ shifts toward *c* axis at higher field. The resistance drops described here are consistent with the MIT that occurs more easily for *H // ab* than *H // c*. The AMR in this region is dramatically enhanced, which amounts 40000%. For *H* > 29.5 kOe, the magnetoresistance change is greatly suppressed, and a small dip is formed near *H // c*, which finally evolves into minimum resistivity at higher fields. The polar plot of the AMR presented in Figure 2c clearly shows uniaxial symmetry, but the maxima and minima positions evolve with field. For small fields such as 10 kOe, the resistivity maxima appears at *H // c* while minima occurs at *H // ab*. As *H* is promoted to ~ 28 kOe, the maximum resistivity at *H // c* is split into two peaks which shift towards *H // ab* with field. Accordingly, the AMR distorts from a dumbbell-like shape to a butterfly-like shape (Figure 2c).

Noteworthy, the AMR above $T_N$ behaviors differently from that below it. The first evident difference is the sign change. As is presented in Figure 2b of the AMR curves at 15 K, the resistivity maxima occur at *H // ab* and minima at *H // c*, which are opposite to that at 2 K. At *H* = 24 kOe, the resistivity for *H // c* (~ 0.59 Ω·cm) is about one order of magnitude smaller than that for *H // ab* (~ 5.8 Ω·cm). The sign change of AMR is consistent with the reversal of the resistivity anisotropy presented in Figure 1d. The second difference is the absence of change in the positions for resistivity maxima and minima when field is changed. The polar plot of AMR remains dumbbell-like shape and no splitting is observed at higher magnetic field (Figure 2d). The difference in the evolution of resistivity maxima and minima for AMR below and above $T_N$ can be understood by considering the angular dependence of magnetization due to interplay between magnetic anisotropy and Zeeman energy.[19]

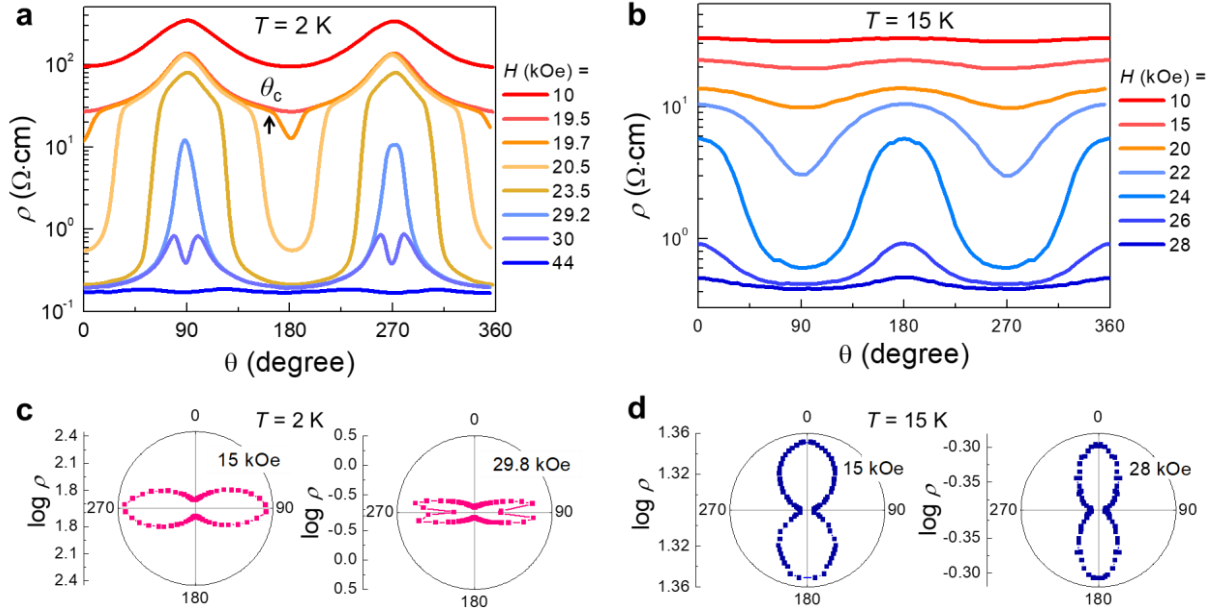

**Figure 2.** Characterization of the anisotropic magnetoresistance. a,b) AMR for different values of $H$ as a function of $\theta$ at 2 K (a) and 15 K (b). The critical angle ($\theta_c$) that corresponds to an additional resistivity drop is indicated by an arrow. c,d) Polar figure of the AMR curves at typical magnetic fields at 2 K (c) and 15 K (d), respectively. For $T = 15$ K, the data with 180° < $\theta$ < 360° are symmetrized from data with 0° < $\theta$ < 180° except for the $H$ = 20 and 22 kOe curves.

The temperature and field dependence of AMR is summarized in a phase diagram shown in **Figure 3**. In order to have a more straightforward comparison between the AMR above and below $T_N$, the AMR is defined as: $(\rho_{H // c} - \rho_{H // ab}) / \rho_{min}$, where $\rho_{min} = \rho_{H // ab}$ below $T_N$ and it becomes $\rho_{H // c}$ above $T_N$. The remarkable AMR arises from the magnetic field direction dependent MIT, which has different critical magnetic field for $H // c$ ($H_{cr}(c)$) and $H // ab$ ($H_{cr}(ab)$), respectively. The AMR exhibits nonmonotonic field and temperature dependence with the following unique characteristics. First, the AMR runs up to several hundredfold, which is exceptionally high in magnetic materials. Second, the AMR is positive in the AFM state and becomes negative in the PM phase. Near the PM-AFM phase transition temperature it is suppressed. Third, the AMR is considerably large even when $T$ is much higher than $T_N$. For example, an AMR of -180% can still be observed at 25 K (~ 2.2 $T_N$).

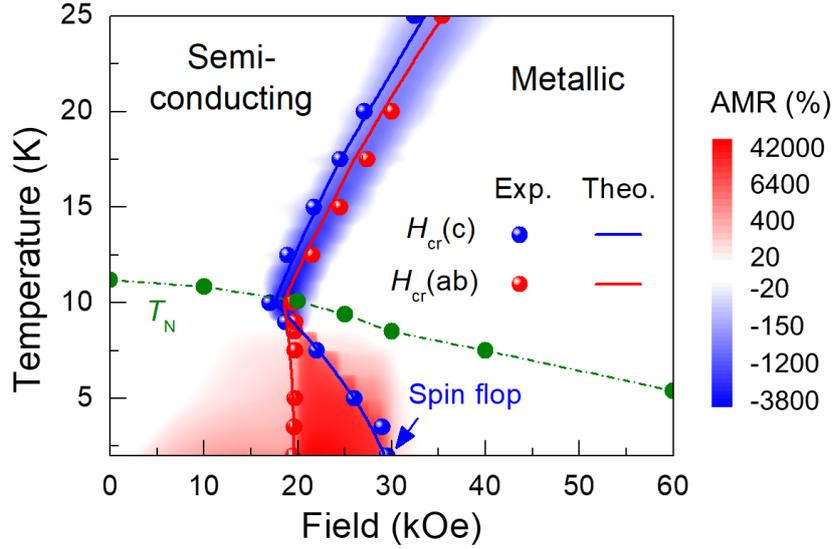

**Figure 3.** Phase diagram of AMR and the MIT. A contour plot of the *T*- and *H*-dependence of the AMR. The experimental and theoretical critical field strength for MIT with *H* // *c* (blue) and *H* // *ab* (red) are also plotted. $T_N$ denotes the AFM to PM transition temperature. The spin-flop field at 2 K is indicated by the green arrow.

The unique vector field dependent MIT and characters of AMR suggest different a mechanism than well-established Lorentz force or spin-orbit coupling picture.[20-22] Although the MIT for *H* // *c* coincides with the spin-flop transition, the absent specific heat capacity change during MIT for *H* // *c* (**Figure S5**) indicates it is coupled with neither structural nor magnetic phase transition. Thus, the colossal AMR presented in Figure 2 are not essentially correlated with magnetic phase transition. This fact is in strong contrast to many phase separation materials and other antiferromagnets of such as $Sr_2IrO_4$.[23-25] Despite of the absence of magnetic phase transition, we found a dramatic change of the Hall resistivity accompanying the MIT. The Hall effect measurement suggests a significant increment of the carrier density by more than one order of magnitude (**Figure S7**), implying a notable change of electronic band structure. The colossal AMR revealed here is thus clearly associated with a vector field dependent MIT and band structure.

In order to interpret the magneto-transport as well as its anisotropy in $EuTe_2$, we investigated the band structure upon magnetic field by performing density functional theory (DFT) calculations and Monte Carlo (MC) simulations. The DFT calculations indicate that the ground state is antiferromagnetic with easy axis along *c* direction.[19] The magnetism is contributed by Eu-*f* electrons with a magnetic moment of 6.95 μB per Eu. The band structures of EuTe$_2$ with AFM state as shown in **Figure 4a,c** indicate that it is an insulator with an indirect band gap of 23.3 meV. With regard to antiferromagnetic spin structure, although time-reversal (T) symmetry and inversion (P) symmetry are both broken, their combination, namely the space-time inversion symmetry (PT), is conserved. Consequently, each band acquires a twofold Kramers degeneracy at each momentum from PT symmetry. Specifically for our material, the bands near the Fermi level are contributed by nonmagnetic itinerant Te-*p* electrons, which hybrid with lower localized Eu-*f* bands. Therefore, the AFM spin structure enables the itinerant Te-*p* electrons to preserve the PT symmetry with the twofold degenerate bands. When the collinear spin array is locally perturbed, the strict PT symmetry is slightly violated, which then splits the doubly-degenerate bands into two singly-degenerate bands. The splitting energy (ΔE) can be expressed as:

$$\Delta E = \frac{M}{M_S}\left(\sqrt{1+\left(\frac{2h}{\Delta}\right)^2}-1\right), \tag{1}$$

where Δ is band interval between local and itinerant orbitals, *h* is a hopping term, *M* and *M$_s$* are the magnetic moment and saturated magnetic moment, respectively. A larger $\frac{h}{\Delta}$ corresponds to stronger hybridization, which along with larger $\frac{M}{M_S}$ could result in an enhanced splitting. Taking Δ = 1.5 eV, *h* = 0.5 eV, and $\frac{M}{M_S}$ = 1 as example, a splitting energy of ~100 meV is obtained, which is two orders of magnitude larger than Zeeman splitting (~1 meV). As a result of band splitting from the slightly canted local spin of Eu (e.g., by ~10°), the conduction band minimum along M-Γ direction and valence band maximum at Γ point are shifted down and up, respectively, as shown in Figure 4b,d. Consequently, the band gap is closed and MIT is induced.

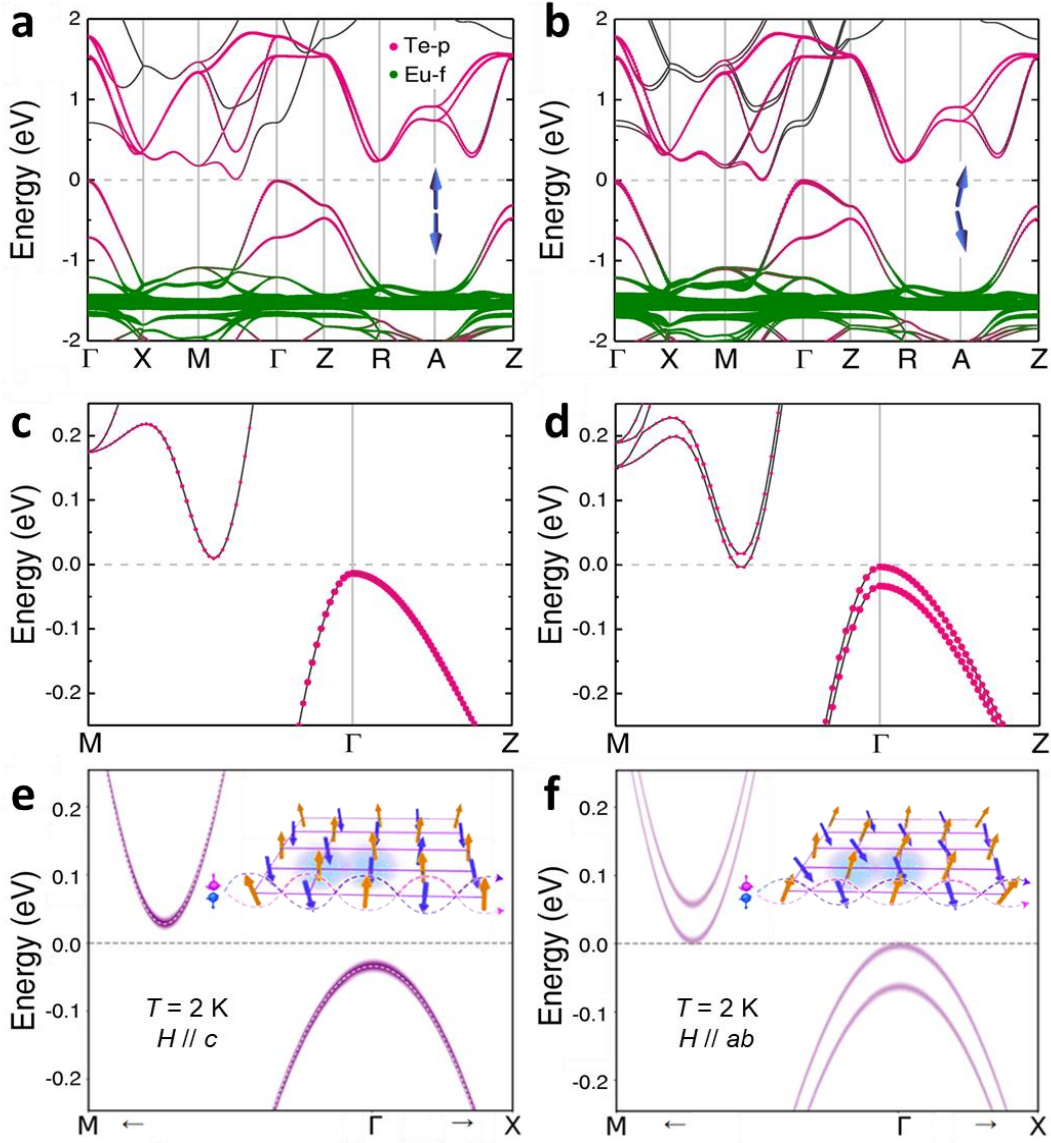

**Figure 4.** Calculated band structures of EuTe$_2$ at different conditions. a,b) Orbital-projected band structure of EuTe$_2$ with AFM and canted spin configurations, respectively. The spin configurations are demonstrated by blue arrows. c,d) Zooming of the bands near Fermi level in (a) and (b), respectively. The non-collinear magnetic order is used in band calculations while spin-orbit coupling is not included. e,f) The band structures simulated with TB model combined with band unfolding technique, based on the spin configurations simulated with MC method. The temperature and magnetic field used in the simulations are 2 K and 25 kOe, respectively. White dashed line in (e) indicates the simulated band structure of primitive cell. Insets illustrate the spin textures response to *H //c* and *H //ab* below the Néel temperature. The blue and orange arrows represent local spin with different directions. The pink and blue dashed lines denote the local spin contributed potential barrier for itinerant electron.

The DFT calculations indicate that the band structure of EuTe$_2$ can be tuned by tilting the spins directions, which in reality can be realized by applying external fields at a specific temperature. At zero temperature and relatively small field (thus no spin slop or spin flip occurs), the AFM structure remains unchanged for *H // c*. Different from *H // c*, spin-canting can occur for *H // ab*. At a finite temperature, thermal energy sets in and forces the spins to fluctuate around their principal axis. The spin configurations with *H // c* and *H // ab* at a finite temperature of 2 K are schematically depicted in the inset of Figure 4e,f, respectively. Note that DFT is not applicable to calculate the band structure under a field at a finite temperature, we thus developed a method to simulate the band structure at experimental compatible condition (*H*, *T*) by (1) simulating the spin configurations in a supercell at a specific field and temperature using MC method, and (2) constructing tight binding model to obtain the corresponding supercell band structure which is then unfolded to Brillouin zone of primary unit cell through a band unfolding technique.[19,26] The obtained band structures at 2 K are shown in Figure 4e,f, which correspond to the spin configurations shown in the insets. The simulated band structures are very similar to our DFT calculation as shown in Figure 4c,d. The main difference is that the simulated one is slightly broadened, which is due to thermal-induced spin fluctuations. The external fields and temperatures play a role like the self-energies in correlated metals and Mott insulators.[27-29] We computed the effective self-energies, and revealed that the spectral function of quasiparticle is mainly distributed below $E_F$ at the experimental condition of 2 K and $H_c$ = 25 kOe, indicating it is in an insulating-like state. As the field direction is changed to *H // ab*, the spectral functions were partially shifted above $E_F$, leading to a metallic state.[19] These results indicate that changing the directions of magnetic field would produce different spin configurations, which then result in a varied band structure of particular near the Fermi energy. As a result, a colossal AMR is induced.

According to equation (1), the band splitting scales with the induced net magnetization moment, which can help to understand the inversion of the magnetotransport anisotropy across $T_N$. As can be seen from Figure S3, at low temperature such as 2 K, the field-induced magnetization for $H // ab$ is larger than that with $H // c$, thus the effective band gap is smaller with $H // ab$, indicating a less resistive state with $H // ab$. However, at higher temperature of roughly $T > T_N$, the field-induced magnetization (effective band gap) for $H // ab$ is smaller (larger) than that with $H // c$, which gives rise to an opposite conduction state as compared to that at low temperature and explains the experimentally observed sign change of AMR. Moreover, a phase diagram of MIT with respect to temperature and magnetic field is constructed according to the combined MC and band unfolding method and is depicted in Figure 3, showing good agreement with experimental results.

**Table 1** lists the AMR in some typical material systems. In nonmagnetic (NM) materials, the Lorentz force induced deviation of carrier trajectories depends on the field direction, leading to an apparent AMR.[20] In materials with high carrier-mobility and anisotropic Fermi surface, the Lorentz force could result in very large magneto-transport anisotropy and thereby a giant apparent AMR of as large as hundreds-fold.[30,31] In contrast, the AMR in magnetic materials mainly arises from anisotropic spin-dependent scattering due to spin-orbit coupling,[9,11] which usually gives rise to a small AMR.[8,32-34] In some materials, large AMR could be induced during magnetic transition upon magnetic field, such as in Mott materials with MIT and antiferromagnetic oxides with spin-flop transition.[24,35] Here the AMR in EuTe$_2$ has fundamentally different mechanism where the vector-field modifies the spin directions to induce distinct varying electronic band structures, a similar philosophy to our previous reported spin-textured band effect.[36,37] The band gap change is much more significant than that in topological materials where only shift of topological points is proposed,[34,38,39] and the AMR exhibits a remarkably large ratio up to 40000%, which is a recording highest value among magnetic materials to the best of our knowledge.

**Table 1. List of AMR and the main mechanisms in selected materials.**

|     | **Materials** | **Main mechanisms** | **AMR value** | **Ref.** |
| --- | --- | --- | --- | --- |
| NM | Cu | Fermi surface anisotropy | 800% @ 4.2 K | [40] |
|    | Bi |  | 1000% @ 2 K | [41] |
|    | Graphene |  | 100% @10K | [42] |
|    | $WTe_2$ |  | 100000% @ 2 K | [30] |
| FM | $Ni_{80}Fe_{20}$ | Spin-orbit coupling | 16% @ 4.2 K | [32] |
|    | $La_{0.7}Ca_{0.3}MnO_3$ | Electron-correlated magnetic phase transition | 600% @ 220 K | [35] |
| AFM | FeRh | Spin-orbit coupling | 1% @ 200 K | [8] |
|     | CuMnAs |  | 0.3% @ 2 K | [14] |
|     | $Sr_2IrO_4$ | Spin-flop transition | 160% @ 35 K | [24] |
|     | $CeAlGe_{0.72}Si_{0.28}$ | Magnetic-coupled topological points | 5% @ 2 K | [38] |
|     | $EuMnSb_2$ |  | 5400% @ 2 K | [39] |
|     | **$EuTe_2$** | **Spin-canting induced band reconstruction** | **40000% @ 2 K** | **This work** |

In particular, the spin-canting induced band reconstruction in AFM can occur without spin-orbit coupling,[19] which inspires the search for functional AFM materials without necessarily incorporating heavy elements. From our results, prominent anisotropic magneto-transport behavior could be expected in antiferromagnets with: (1) a narrow band gap, so that the magnetic-field-induced band splitting could lead to considerable band change; (2) a large localized spin moment, so that it could response easily to external magnetic field; (3) strong hybridization between orbitals of localized spin and itinerant carriers. These requirements would be fulfilled in compounds containing rare-earth elements (large local spin moment) and non-metal elements from IV to VI group (possible narrow band gap).

To summarize, our results show that due to a strong hybridization between orbitals of localized Eu spin and Te itinerant carriers, the band structure of antiferromagnetic rare-earth dichalcogenide EuTe$_2$ is strongly modified by the local configuration of spin direction, which leads to gap-close when field is switched from *c* axis to *ab* plane and thus MIT. As a result of such vector field dependent band structure, a colossal AMR of 40000% has been revealed in this new class of antiferromagnets, which affords unprecedented opportunity for AFM spintronics application. Although in the present case the spin-canting is driven by a magnetic field, it could also be realized by other ways such as the Dzyaloshinskii-Moriya (DM) interaction that favors non-colinear spin structure.[43,44] Thus our findings suggest a way towards novel antiferromagnetic spintronic devices, where different resistance states are realized by controlling the DM interaction.

**Experimental Section**

*Crystal growth and structural characterization*: EuTe$_2$ single crystals were grown by self-flux methods. Eu (99.9%) and Te (99.999%) shots were combined in the molar ratio of 1: 10 and then sealed in an evacuated quartz ampoule. Afterwards, the ampoule was put in a muffle furnace and slowly heated to 850 °C in 100 hours and held for 3 days, then slowly cooled to 450 °C in 300 hours followed by centrifuging at this temperature to separate crystals from the Te flux. Crystals with dark and mirrorlike surface were obtained and used in our experiments. The typical size of the single crystals is 4 mm × 1.5 mm × 1 mm. X-ray diffractions were measured on a Bruker D8 Discovery single crystalline X-ray diffractometer at room temperature. The X-ray wavelength is 1.5405 Å. Theta-omega scans along different Bragg diffraction planes were performed.

*Magnetic and transport property measurements*: Magnetic measurements were conducted on a superconducting quantum interference device magnetometer (SQUID, Quantum Design). Typical cooling and warming rates are 5 K per minute. In order to remove the remnant magnetic field of superconducting coil, before loading the sample, a field of 30 kOe was applied and then decreased to zero in oscillation mode at room temperature. Isothermal field dependent magnetizations for *H // c* and *H // ab* were measured at temperatures between 2 K and 20 K. The magneto-transport measurements were carried out using a Physical Property Measurement System (PPMS, Quantum Design) equipped with a motorized sample rotator. The typical cooling rate is 2 K per minute, and the warming rate is 3 K per minute. A standard four-probe method was used for the resistivity measurements. Conductive silver paste was used to connect the Pt wires to the sample surface. The Hall effect was measured using four terminal Hall bar configurations with *I // c* and *H // ab*. The magnetic fields were swept from -$H_{max}$ to $H_{max}$, and the influence of misalignment between the Hall electrodes on Hall signal was eliminated by subtracting the negative field branch from the positive field branch.

*Heat capacity measurement:* The specific heat capacity was measured by a time-relaxation method in PPMS. During measurement, the chamber was evacuated to a pressure of better than $1 \times 10^{-5}$ Torr. The temperature dependent specific heat capacity was measured during the warming process, and the isothermal magnetic field dependent specific heat capacity was measured in the field ascending case. The magnetic field was applied in the *ab* plane.

*Monte Carlo simulations*: Monte Carlo simulation was performed based on Metropolis algorithm. A spin Hamiltonian of $H = \sum_{\langle i,j \rangle} J_{\perp} \left( S_i^x S_j^x + S_i^y S_j^y \right) + \sum_{\langle i,j \rangle} J_z S_i^y S_j^y - \sum_i D \left( S_i^z \right)^2 - \sum_i \mu_B g \vec{B} \cdot \vec{S}_i$ was adopted for the simulation, where the four terms are the in-plane magnetic coupling, the out-of-plane magnetic coupling, the single ion magnetic anisotropy, and Zeeman coupling, respectively.

*First-principles calculations:* The DFT calculations were performed within the generalized gradient approximation and the projector augmented wave method as implemented in the Vienna ab initio simulation package (VASP). Valence electron configurations of $5s^2 6p^6 4f^7 6s^2$ and $5s^2 5p^4$ in calculations were considered for Eu and Te, respectively. The PBEsol functional was adopted for structure relaxation. The non-collinear magnetic scheme was used in spin canted calculations, and spin-orbit coupling (SOC) scheme was also considered for comparison. The band structures for AFM and FM with modified Becke-Johnson (MBJ) exchange potential and Heyd-Scuseria-Ernzerhof (HSE06) hybrid density functional has been performed to confirm the bands gap variation between AFM and FM states.

Band unfolding technique was used to obtain the simulated bands. A super cell (SC) tight-binding (TB) Hamiltonian:

$$H = t_1 \sum_{\langle i,j \rangle} \sum_{\gamma\alpha} C^\dagger_{i\alpha\gamma} C_{j\alpha\gamma} + t_2 \sum_{\langle\langle i,j \rangle\rangle} \sum_{\gamma\alpha} C^\dagger_{i\alpha\gamma} C_{j\alpha\gamma} + t_3 \sum_{\{\{i,j\}\}} \sum_{\gamma\alpha} C^\dagger_{i\alpha\gamma} C_{j\alpha\gamma} + \lambda \sum_{i\alpha\gamma\gamma'} C^\dagger_{i\alpha\gamma} C_{j\alpha\gamma'} S_{i\alpha} \cdot \sigma_{\gamma\gamma'}$$

was constructed to input spin configuration obtained from MC simulation, where the four terms are nearest-neighbour hopping terms, next nearest-neighbour, third nearest-neighbour hopping and exchange coupling between localized electron spin and itinerant electron respectively.

**Acknowledgements**
We acknowledge the financial support from National Nature Science Foundation of China (51525103, 51931011, 11774360, 51701231, 11904373, 11874264), National Key R&D Program of China (2017YFA0303602), Key Research Program of Frontier Sciences, Chinese Academy of Sciences (ZDBS-LY-SLH008). Work at USTC is supported by Nature Science Foundation of China (11974325). Work at Sun Yat-Sen University was supported by the National Nature Science Foundation of China (11904414), Natural Science Foundation of Guangdong (2018A030313055), National Key R&D Program of China (2019YFA0705700). Huali Yang, Qing Liu, and Zhaoliang Liao contributed equally to this work.

Supporting Information

**Observation of an Unusual Colossal Anisotropic Magnetoresistance Effect in an Antiferromagnetic Semiconductor**

*Huali Yang, Qing Liu, Zhaoliang Liao, Liang Si, Peiheng Jiang, Xiaolei Liu, Yanfeng Guo, Junjie Yin, Meng Wang, Zhigao Sheng, Yuxin Zhao, Zhiming Wang, Zhicheng Zhong\*, Run-Wei Li\**

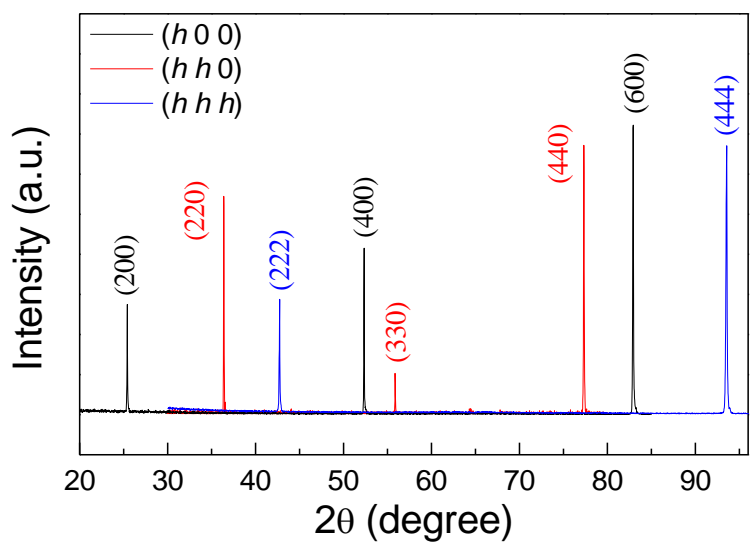

**Figure S1**. X-ray diffraction patterns. Diffraction patterns obtained from the same crystal and with different Bragg diffraction conditions. The diffraction peaks are indexed as (*h* 0 0), (*h h* 0), and (*h h h*) planes, respectively. According to Bragg equation, lattice parameters were determined from diffraction peaks to be *a* = *b* = 6.980 Å and *c* = 8.376 Å.

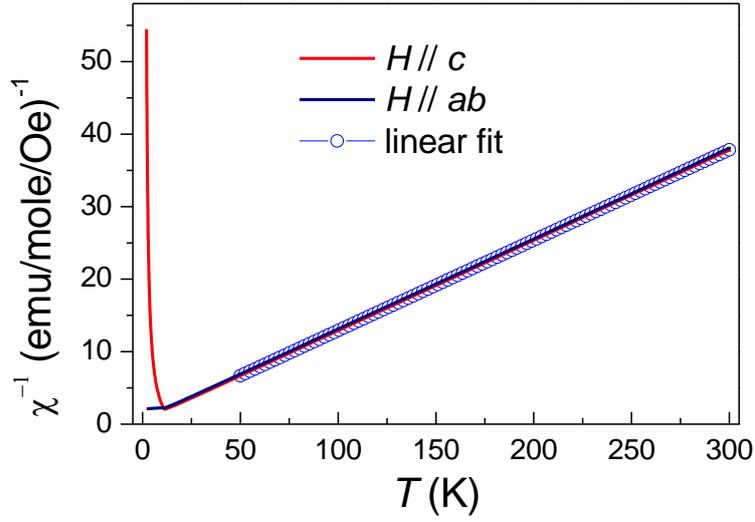

**Figure S2**. Fitting of the magnetic susceptibility. The reciprocal of magnetic susceptibility as a function of temperature for $H // c$ and $H // ab$, respectively. According to the Curie-Wiess formula of $\chi^{-1} = (T - \theta) / C$, both the Curie-Wiess temperature $\theta$ and the Curie-Wiess constant $C$ can be obtained from a linear fit of $\chi^{-1}(T)$. The fitted results are $\theta$ = -5.36 K and $C$ = 8.029 emu·mole$^{-1}$·Oe$^{-1}$·K. The negative $\theta$ indicates antiferromagnetic interaction below $T_N$. According to the formula $C = \dfrac{N_A}{3k_B}\mu_{eff}^2$, where $N_A = 6.022 \times 10^{23}$ is the Avogadro constant and $k_B = 1.38 \times 10^{-23}$ J / K is the Boltzmann constant, we obtained the effective magnetic moment ($\mu_{eff}$) to be 7.874 $\mu_B$. The effective moment is close to the theoretical expectation of $\mu = g\sqrt{S(S+1)}\mu_B$ = 7.9 $\mu_B$ with $S = \dfrac{7}{2}$ for Eu$^{2+}$, where g = 2 is the spin Landau g factor of electrons.

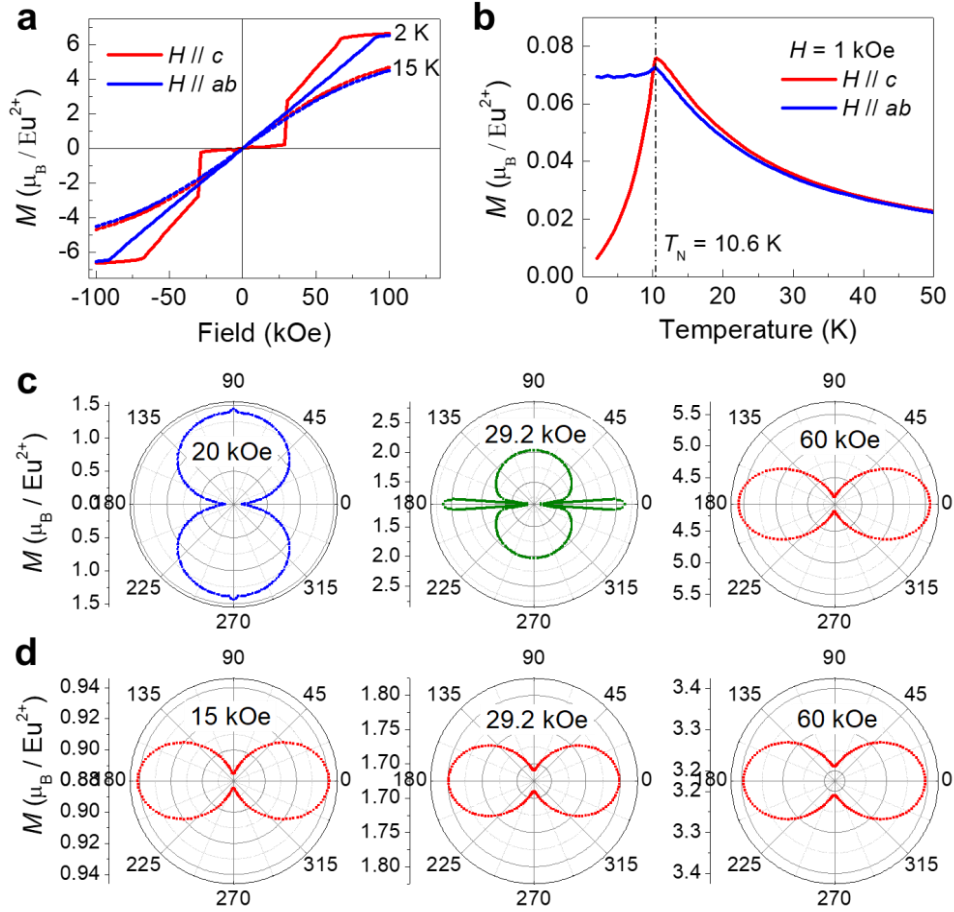

**Figure S3**. Simulated magnetization response at various external conditions. a) Simulated *M-H* curves at 2 K and 15 K. b) Simulated *M-T* curves at 1 kOe for *H // c* and *H // ab*, respectively. c,d) Polar figure of magnetic field direction depended magnetization obtained from Monte Carlo simulations at *T* = 2 K (c) and *T* = 15 K (d). Due to symmetry of system the data in the range of 90° - 360° are reproduced from that in the range of 0° - 90°. 0° correspond to *H // c* and 90° correspond to *H // ab*.

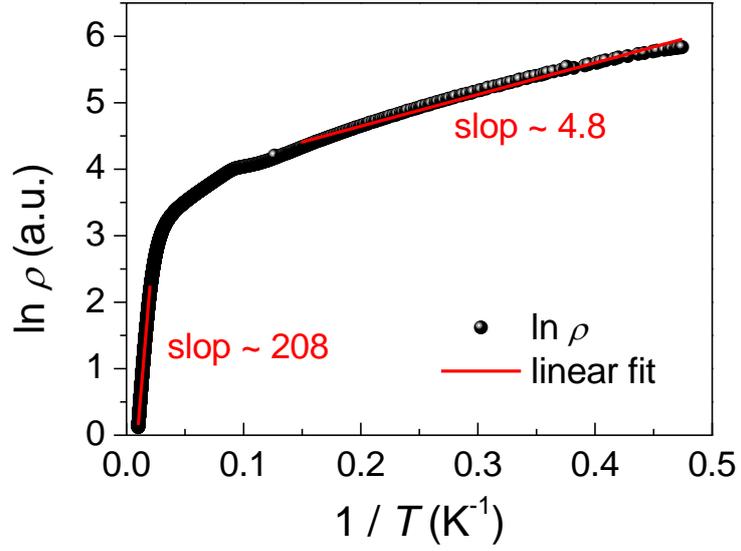

**Figure S4**. Estimation of the thermal activation gap. *Ln*-scale resistivity is plotted as a function of the reciprocal of temperature. The red lines correspond to linear fit in the region above $T_N$ and below $T_N$, respectively. According to the thermal activation model: $\ln \rho = \ln \rho_0 + \frac{E_a}{k_B T}$, where $E_a$ is thermal activation energy, and $k_B$ is the Boltzmann constant, we obtained thermal activation energies above $T_N$ and below $T_N$ of ~18 meV and ~ 0.4 meV, respectively

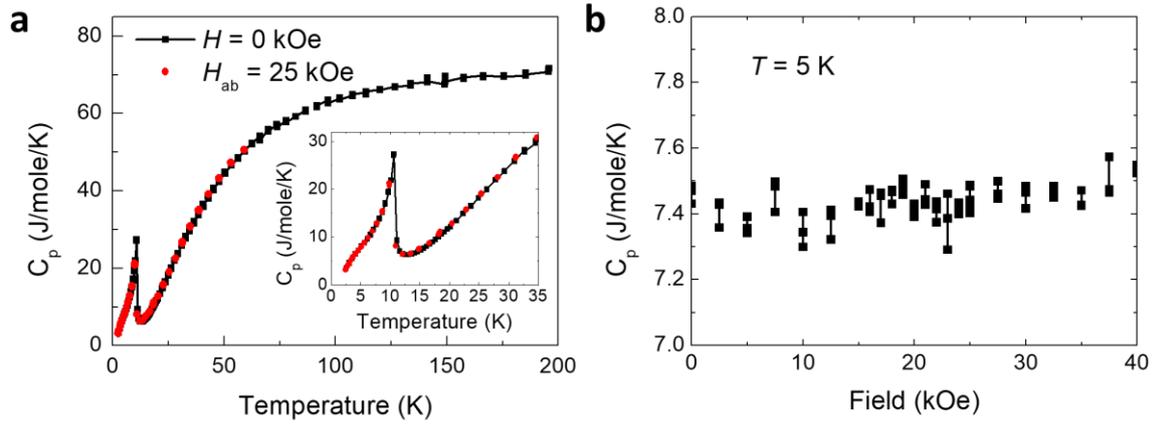

**Figure S5**. Temperature and magnetic field dependence of the specific heat capacity. a) Temperature dependent specific heat capacity at 0 kOe and $H_{ab}$ = 25 kOe, respectively. Inset shows the magnified image below 35 K. No evident change of specific heat capacity has been observed. b) Magnetic field dependent specific heat capacity at $T$ = 5 K. There is also no observable change of specific heat capacity up to 40 kOe, indicating the absence of magnetic phase transition accompanying MIT (which occurs at ~ 20 kOe).

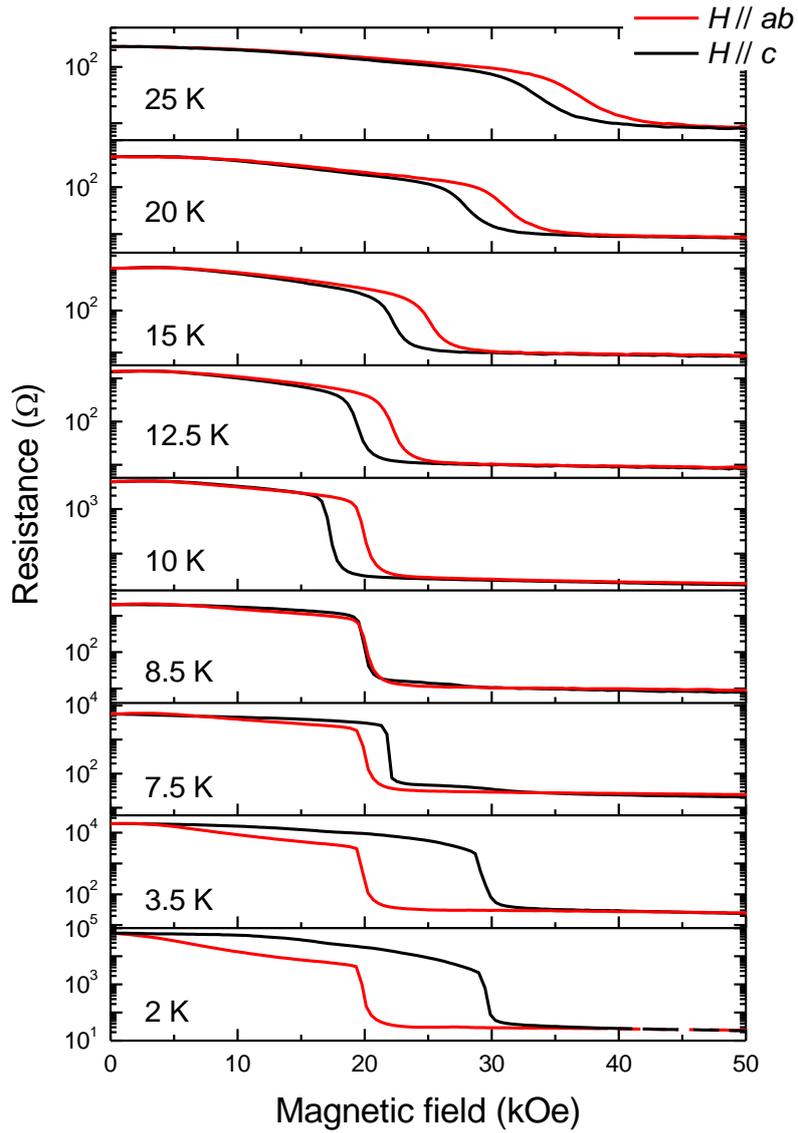

**Figure S6**. Isothermal magnetoresistance of EuTe$_2$. The data was taken at selected temperatures between 2 K and 250 K with $I // c$, and $H$ was applied parallel to $c$-axis and $ab$ plane, respectively.

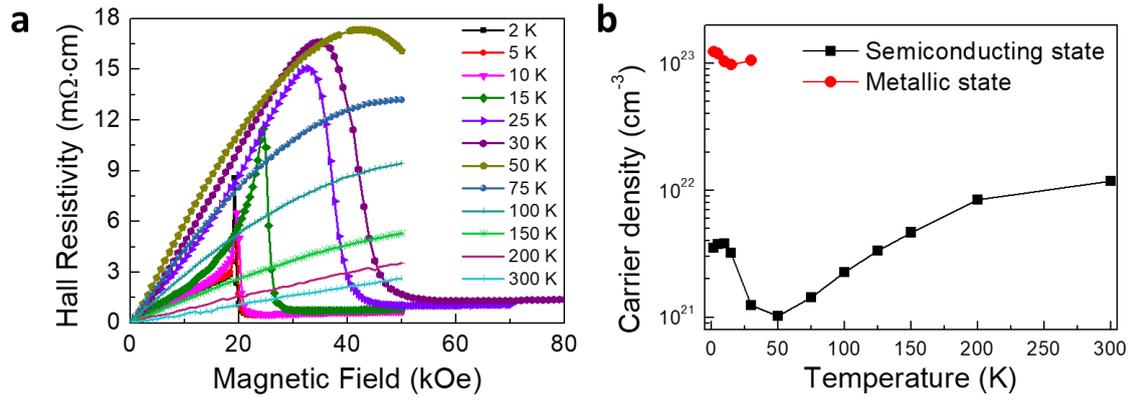

**Figure S7**. Hall measurement results. a) Hall resistivity of EuTe$_2$ measured at various temperatures between 2 K and 300 K. b) The estimated carrier density by using a single band model.

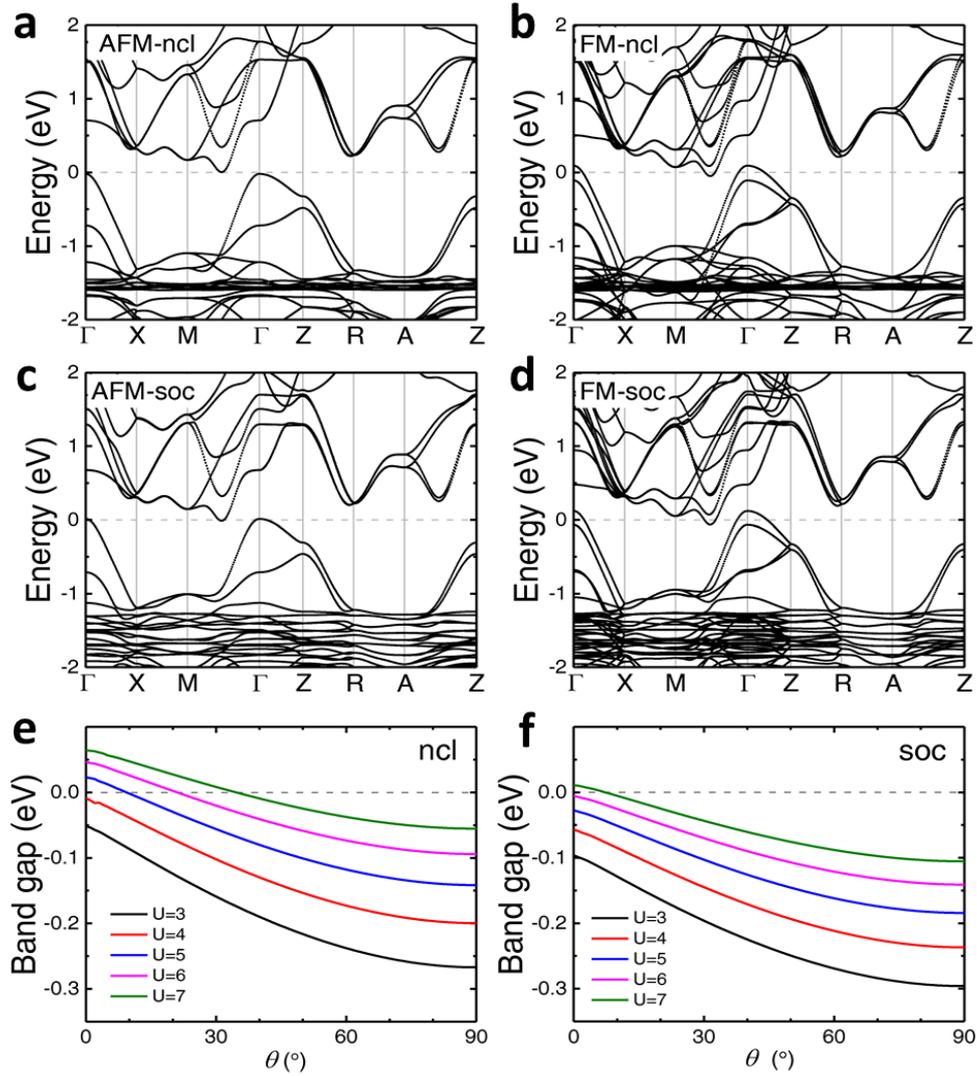

**Figure S8**. DFT-calculated band structures of EuTe$_2$. a-d) Band structures calculated with PBEsol functional. Both AFM and FM orders are calculated by adopting (a,b) non-collinear (ncl) and (c,d) SOC schemes, respectively. e,f) Band gap variation with respect to different spin canted angles ($\theta$) as shown in the inset calculated with PBEsol functional. The $\theta = 0°$ and $\theta = 90°$ cases correspond to AFM and FM orders, respectively.

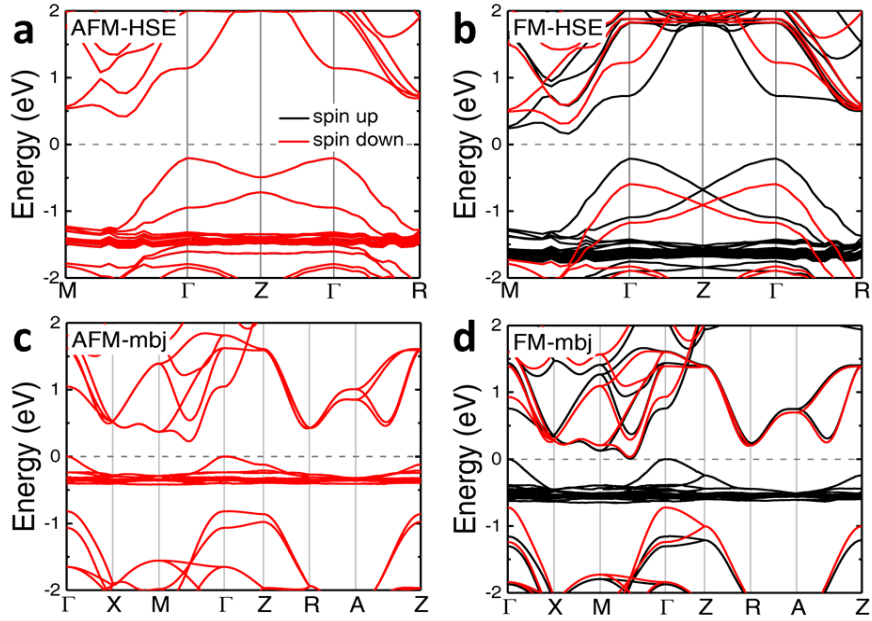

**Figure S9**. Band structures of AFM and FM state. a,b) Calculated with Heyd-Scuseria-Ernzerhof (HSE06) hybrid functional. c,d) Calculated with modified Becke-Johnson (mbj) exchange potential. In (a) and (c), the spin up and spin down channel overlaps with each other.

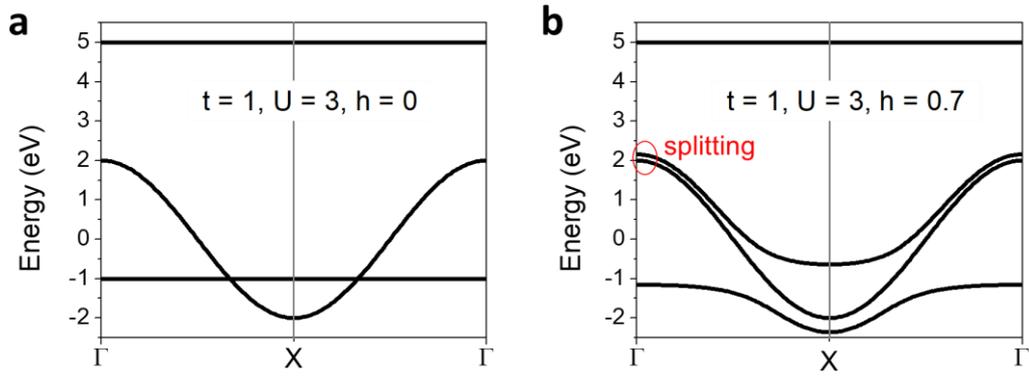

**Figure S10**. Illustration of band structure of hybridization model. a) Band with no hybridization, i.e., $h = 0$. b) Band with $h = 0.7$.

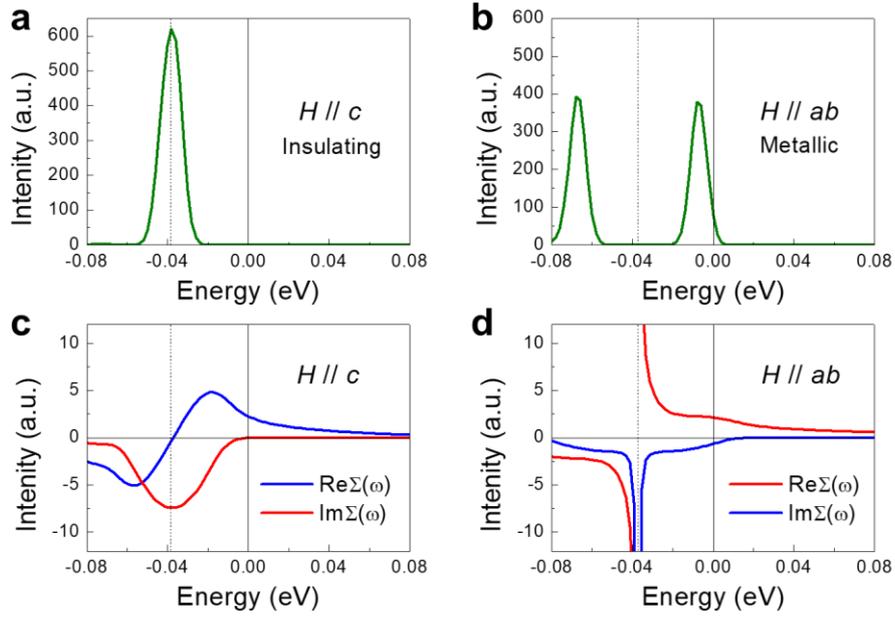

**Figure S11**. Spectral functions and Self-energy at $T$ = 2 K and $H$ = 25 kOe. a,b) Spectral function with $H // c$ (a) and $H // ab$ (b) respectively. c,d) Self-energy with $H // c$ (c) and $H // ab$ (d) respectively. Dashed gray lines mark the calculated valance band maximum energies based on TB model.

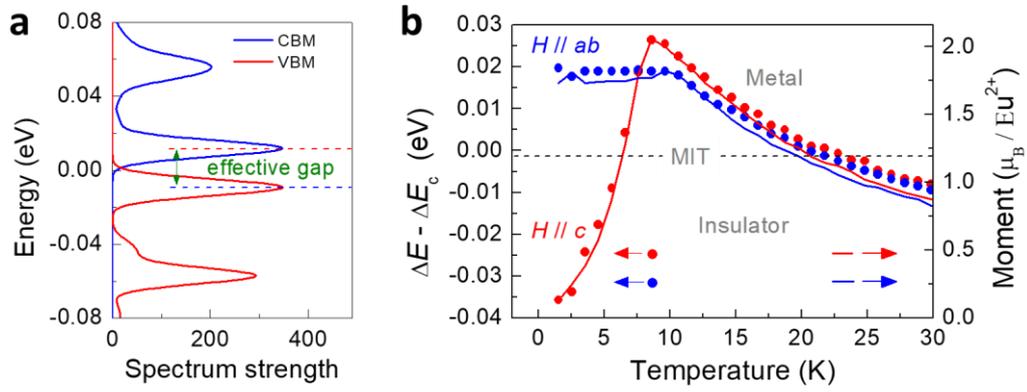

**Figure S12**. Results from band unfolding technique. a) Schematic diagram of defining effective gap. b) Temperature dependent magnetization and the effective gap with $H = 25$ kOe calculated by the peak of spectrum around valance band maximum subtract that in conduction band.

## S1. Monte Carlo simulation

To simulate the magnetization response with respect to different temperature and magnetic field, a classic Monte Carlo method based on Metropolis algorithm was used. Based on the lattice symmetry of EuTe$_2$, and by comparing all possible AFM configurations with experimental data, we finally confirm a spin Hamiltonian in a following form:

$$H=\sum_{\langle i,j \rangle} J_\perp \left(S_i^x S_j^x + S_i^y S_j^y\right) + \sum_{\langle i,j \rangle} J_z S_i^y S_j^y - \sum_i D\left(S_i^z\right)^2 - \sum_i \mu_B g \vec{B} \cdot \vec{S}_i \quad (1)$$

There are four terms including in-plane AFM coupling $J_\perp > 0$, out-of-plane FM coupling $J_z < 0$, single ion magnetic anisotropy $D > 0$, and Zeeman coupling.

In the simulations, we discarded the first 4,000,000 Monte Carlo steps (MCSs) to avoid the effects of initial configurations, and then the subsequent 100,000 MCSs were used for a simulation output. In a 3D lattice with 30 × 30 × 30 atoms, the simulated results are in good agreement with experimental magnetization curves (Figure 1c in main text) with using $J_1$ = 0.442 meV, $J_2$ = -0.754 meV and $D$ = 0.262 meV. In addition, the Zeeman energy term was calculated using a saturated magnetic moment of 7.0 μ$_B$. With these parameters, we also simulated the $M$-$H$ curve at high temperature of $T$ = 15 K as shown in Figure S3a. To estimate the Néel temperature, we use a sizeable lattice with 80 × 80 × 80 atoms in order to reduce the influence of size effect. The simulated Néel temperature is 10.6 K (Figure S3b), which is very close to the experimental value of 11.2 K.

The Monte Carlo simulation has also been applied to reveal the response of magnetization to external field directions. As is shown in Figure S3c,d, at both $T$ = 2 K and 15 K, the angular dependent magnetization curves exhibit uniaxial symmetry, irrespective of the magnetic field strength. At $T$ = 2 K, with the increase of magnetic field strength, the position with maximum magnetization evolves from $H$ // $ab$ to $H$ // $c$. Whereas at $T$ = 15 K, position with maximum magnetization is always along $c$ axis, irrespective of the field strength. These phenomena are consistent with the AMR behavior (Figure 2 in the main text), and their link can be well explained based on STB effect.

**S2. First principle calculation**

We performed the DFT calculations within the generalized gradient approximation and the projector augmented wave method as implemented in the Vienna ab initio simulation package (VASP).[45-47] Valence electron configurations of $5s^26p^64f^76s^2$ and $5s^25p^4$ in calculations were considered for Eu and Te, respectively. The PBEsol functional was adopted for structure relaxation.[48] The relaxed lattice constants are $a = b = 6.97$ Å and $c = 8.18$ Å. The Brillouin zone was sampled with $8 \times 8 \times 7$ $k$ mesh and the kinetic energy cutoff was set to 500 eV. The non-collinear magnetic scheme was used in spin canted calculations, and spin-orbit coupling (SOC) scheme was also considered for comparison. Magnetic calculations indicate that the ground state could be either A-type AFM or C-type AFM configurations since they differ only 1.6 meV/Eu in energy. Further calculations demonstrate that different AFM configurations have negligible influence on the band structure. Therefore, we adopted A-type AFM configurations in band structure calculations. The band structures for AFM and FM with modified Becke-Johnson (MBJ) exchange potential and Heyd-Scuseria-Ernzerhof (HSE06) hybrid density functional has been performed to confirm the bands gap variation between AFM and FM states, both give qualitatively similar results (see in Figure S9).

## S3. Toy model for hybridization between itinerant and localized electrons

In this part, we use TB toy model to illustrate the effect of hybridization between itinerant and localized electrons and the band structure variation induced by the hybridization. By considering two kind of orbitals (itinerant and localized orbitals) with spins (↑ and ↓), i.e., $\phi_{L\uparrow}$, $\phi_{L\downarrow}$, $\phi_{I\uparrow}$, $\phi_{I\downarrow}$, the hamiltonian is written as:

$$H = \begin{pmatrix} U+2 & 0 & 0 & 0 \\ 0 & -U+2 & 0 & h \\ 0 & 0 & 2t\cos(k) & 0 \\ 0 & h & 0 & 2t\cos(k) \end{pmatrix} \quad (2)$$

Where $U$, $h$, and $t$ denote Coulomb repulsion, orbital hybridization and nearest hopping, respectively, and $k$ is wavevector. The band dispersion of itinerant electron is changed by the hybridization which causes two-fold degenerate band to split (Figure S10). The splitting at specific k points can be expressed as $\frac{M}{M_S}\left(\sqrt{1+\left(\frac{2h}{\Delta}\right)^2}-1\right)$, where $\Delta$ is band interval between localized and itinerant orbitals, $h$ is onsite hopping coefficient between them, $M$ and $M_s$ are the magnetic moment and saturated magnetic moment, respectively. These facts indicate that a strong STB effect might occur in a system where the spin is strongly coupled to the itinerant electrons.

## S4. Band simulation

To obtain the STB under different magnetic field and temperature, we constructed a large supercell (SC) to model complex spin configurations, and calculate the corresponding band structures, and then unfold the SC band to the BZ of primitive cell (PC) to obtain the band variation information by comparing with PC band.

The spin configuration in SC was obtained using Monte Carlo simulation as discussed above. Insets of Figure 4e,f illustrate spin configurations at low temperature with $H // c$ and $H // ab$ respectively. With the obtained spin configurations, we calculated the band structure using a tight banding (TB) toy model with following Hamiltonian:

$$H = t_1 \sum_{\langle i,j \rangle} \sum_{\gamma\alpha} C^\dagger_{i\alpha\gamma} C_{j\alpha\gamma} + t_2 \sum_{\langle\langle i,j \rangle\rangle} \sum_{\gamma\alpha} C^\dagger_{i\alpha\gamma} C_{j\alpha\gamma} + t_3 \sum_{\{\{i,j\}\}} \sum_{\gamma\alpha} C^\dagger_{i\alpha\gamma} C_{j\alpha\gamma} + \lambda \sum_{i\alpha\gamma\gamma'} C^\dagger_{i\alpha\gamma} C_{j\alpha\gamma'} S_{i\alpha} \cdot \sigma_{\gamma\gamma'} \quad (3)$$

Where $i$ and $j$ represent lattice vector coordinates, $\alpha$ and $\gamma$ label orbital and spin index, $t_1$, $t_2$, $t_3$ are the nearest-neighbour hopping terms, next nearest-neighbour and third nearest-neighbour hopping constant, respectively. In the fourth term, $\lambda$ is exchange coupling between localized electron spin and itinerant electron spin; $S_i$ and $\sigma$ label localized electron spin and itinerant electron spin, respectively. Note that SOC is not included in our model since SOC does not essentially change the results (Figure S8). In this toy model, each atom includes 2 orbitals with two onsite potential energy of $\varepsilon_1$ and $\varepsilon_2$, respectively. The parameters $t_1 = 0.8$ eV, $t_2 = 0.08$ eV, $t_3 = 0.16$ eV, $\lambda = 0.12$ eV, $\varepsilon_1 = -3.224$ eV, $\varepsilon_2 = 0.906$ eV were used in simulations.

For comparison, we also calculated the band structure based on PC, where a super-AFM cell was used in TB model with lattice vector $a_1 = (1, -1)$ and $a_2 = (1, 1)$. Each unit cell has two atoms with opposite local spin direction and they are located at (0, 0) and (0.5, 0.5) in direct coordinate system, respectively. The simulated band structure of primitive cell is indicated by white dashed lines in Figure 4e in main text. In the unfolding procedure,[49] we calculated the spectral function A($w$,$k$) expressed by the eigen-function of the PC system,

$$A(k,w) = \sum_i \left\langle \psi^{PC}_{ik} \left| -\frac{1}{\pi} Im\hat{G}(\omega + i0^+) \right| \psi^{PC}_{ik} \right\rangle \quad (4)$$

where $w$ is energy and $k$ is wave vector. $\psi^{PC}_{ik}$ is the eigen-function of PC system and $i$ is the band index. G is the Green's function of SC system and $0^+$ represent the positive infinitesimal. Hence the obtained unfolding bands in the PC BZ with their particular spectral weights represent the contribution from the eigenstates of the PC system to the energy spectrum of the target SC system.[50] Since band unfolding based on 3D lattice has vast atoms and thus requires a too heavy computation. We construct a 2D supercell TB model for all analysis, whose spin configurations are extracted from single layer atoms of 3D lattice. In order to reflect the effect of all atom spin configurations in 3D lattice, the spectral function corresponding to every layer of the spin configurations was calculated separately and added together. A 2D lattice with 30 × 30 atoms TB model was constructed here and we used the same parameters as that in PC case. We input the above obtained spin configuration into supercell TB model and unfolded it. By averaging the sum of spectral function, we obtained the corresponding unfolded band. Since it is hard to visually distinguish the gap state from the spectrum, we plot the spectral function at special BZ $k$-point A($w$) in the vicinity of Fermi level, which provides an estimation of band gap for the system.

## S5. Self-energy analysis

Physically, any band structure's modifications can be attributed to the effects from self-energy $\Sigma(\omega)$. The real part $\text{Re}\,\Sigma(\omega)$ shifts the quasi-particle bands while the imaginary part $\text{Im}\,\Sigma(\omega)$ induces broadening effects. One example is the diverging real part of self-energy in $V_2O_3$ which drives a Mott-insulator transition when the correlation effects become stronger. [51,52]

We applied Anderson Impurity Model to solve the real frequency self-energy of $EuTe_2$ and explain the band spin-splitting as well as the MIT at suitable external conditions (*H*, *T*). To compute the real frequency self-energy, we need to obtain both non-interacting Green's functions *g* and local interacting Green's function *G*, which were computed by non-interacting and local interacting Hamiltonian, respectively. Specially, the simulated tight-binding Hamiltonian is adopted as non-interacting Hamiltonian under 0 K and 0 T. The spectral functions $A(\omega)$ from Monte-Carlo simulation are adopted as particle descriptions under finite temperatures and fields.

The non-interacting Green's functions are:

$$g(\omega) = \frac{1}{\omega - \varepsilon_k} \quad (5)$$

$\varepsilon_k$ is the tight-binding Hamiltonian at 0 K and 0 T, here we use local approximation by making k = Γ. As to the interacting Green's functions, we compute them according to the relationship between $A(\omega)$ and the imaginary part of Green's functions:

$$A(\omega) = \left(-\frac{1}{\pi}\right) \text{Im}\, G(\omega) \quad (6)$$

Hence, from the tight-binding spectra and Monte-Carlo spectra, the real and the imagine part of Green's function can be obtained by applying Kramers-Kronig relations. The relationship between non-interacting Green's functions *g*, interacting Green's functions *G* and self-energy $\Sigma(\omega)$ is defined by Dyson equation:

$$g^{-1} = G^{-1} + \Sigma(\omega) \quad (7)$$

The computed spectral functions $A(\omega)$ and self-energy are shown in Figure S11. At k = Γ, the non-interacting Green's functions *g* is described by a $\delta$ function at $\varepsilon$ = -0.036 eV. At finite temperature, this energy level is broadened. We calculated the spectral functions at *T* = 2 K and $H_c$ = 25 kOe, and found that the energy level is in a singlet state and is distributed between -0.06 to -0.02 eV as shown in Figure S11a. The corresponding self-energy $\Sigma(\omega)$ is computed according to Eq. (7), and it is shown in Figure S11c. On the other hand, as the field is applied along *ab* plane, the singlet state splits into two states (Figure S11b). The Monte-Carlo simulated spectral functions exhibit typical metallic behavior: part of the spectral functions of the higher state crosses $E_F$, leading to hole-like conductivity. The corresponding self-energies are shown in Figure S11d. The real-part of $\Sigma(\omega)$ becomes diverging around the tight-binding $\varepsilon$ = -0.036 eV. This splits the state at Γ into two states, i.e., the band-splitting and hole-like conductivity are triggered by the diverging $\text{Re}\,\Sigma(\omega)$. Moreover, the imaginary part of $\Sigma(\omega)$ further broadens these states, leading to the metallic states.

## S6. Estimation of temperature depended band gap

The effect of various spin configurations on band structure at finite temperature was estimated using band-unfolding scheme described in the above section. In order to analyze the MIT at different temperature, we mainly discuss the band gap evolution since it determines the MIT. Here, we define an effective band gap as the energy difference between peak of spectrum around valance band (PVB) maximum and conduction band (PCB) minimum ($\triangle E$) (see in Figure S12a). Figure S12b plots the temperature dependent effective band gap along with the temperature dependent magnetization for a comparison. They have quite similar temperature dependence behavior. Therefore, we can derive a linear correlation between magnetization and effective band gap. We set $H = 19.5$ kOe and $T = 2$ K in accordance to the experimental condition of MIT, and obtained a critical effective gap $\triangle E_C$ corresponding to the MIT. Regarding a $c$-axis parallel field $H_c = 25$ kOe, the effective gap increases with temperature and reaches $\triangle E_C$ at ~ 7 K, resulting in a insulator to metal transition. As the temperature is raised to ~ 22 K, the effective gap is reduced to be lower than $\triangle E_C$, consequently leading to a metal to insulator transition. While for field in $ab$-plane $H_{ab} = 25$ kOe, on the other hand, the effective gap is always higher than $\triangle E_C$ until $T > $ ~ 20 K, marking a lower critical temperature than $H_c = 25$ kOe. These characters agree very well with the experimental results depicted in Figure 1d. By comparing the gap and magnetization curves of $H // c$ with that of $H // ab$, one can conclude that EuTe$_2$ is more metallic for $H // ab$ than $H // c$ below $T_N$. This situation reverses above $T_N$, which explains the experimental observed reversal of AMR.

Due to the correlation between magnetization and effective gap, we first carry out Monte Carlo simulation at different temperature and magnetic field, and then extract MIT critical magnetization at specific $T$ and $H$ in order to plot a phase diagram. It is noteworthy here that a key character of sign change of AMR can be repeated by our theory model based on STB effect, as shown in Figure 3 in main text.